\documentclass[twocolumn,pra,aps]{revtex4}
\usepackage{graphicx, color, fancybox}
\usepackage{amsfonts}
\usepackage{amssymb,amsmath}

\begin{document}
\title{State discrimination with error margin and its locality}
\author{A. Hayashi, T. Hashimoto, and M. Horibe}
\affiliation{Department of Applied Physics\\
           University of Fukui, Fukui 910-8507, Japan}

\begin{abstract}
There are two common settings in a quantum-state discrimination problem. 
One is minimum-error discrimination where a wrong guess (error) is allowed 
and the discrimination success probability is maximized. 
The other is unambiguous discrimination where errors are not allowed but the 
inconclusive result ``I don't know" is possible. We investigate a discrimination 
problem with a finite margin imposed on the error probability. The two common settings 
correspond to the error margins 1 and 0. For arbitrary error margin, we determine 
the optimal discrimination probability for two pure states with equal 
occurrence probabilities. We also consider the case where the states to be 
discriminated are multipartite and show that the optimal discrimination 
probability can be achieved by local operations and classical communication. 
\end{abstract}

\maketitle

\newcommand{\ket}[1]{|\,#1\,\rangle}
\newcommand{\bra}[1]{\langle\,#1\,|}
\newcommand{\braket}[2]{\langle\,#1\,|\,#2\,\rangle}
\newcommand{\mbold}[1]{\mbox{\boldmath $#1$}}
\newcommand{\sbold}[1]{\mbox{\boldmath ${\scriptstyle #1}$}}
\newcommand{\tr}[1]{{\rm tr}\!\left[#1\right]}
\newcommand{\trm}{{\rm tr}}

\section{Introduction}
Suppose we are given an unknown quantum state $\rho$, which is guaranteed to be one in a set of 
known states $\{\rho_a\}$ with some known occurrence probabilities. 
The task of quantum-state discrimination \cite{Chefles00} is to optimally identify the input state $\rho$ 
with one of $\{\rho_a\}$ by performing some measurement on $\rho$. 
Usually, one of two conditions is imposed on the probabilities of making a wrong guess (error). 
In one setting, the discrimination success probability is optimized without imposing any 
condition on error probabilities \cite{Helstrom76}. This is called minimum-error discrimination since 
the mean error probability is minimized as a consequence. 
On the other hand, an incorrect identification (error) is not allowed in unambiguous discrimination 
\cite{Ivanovic87,Dieks88,Peres88,Jaeger95}. 
Instead, the inconclusive result ``I do not know" is possible when one is not 
certain about the identity of the input state.  

Some interesting alternative approaches have been proposed. 
Croke et al. considered a maximum-confidence measurement \cite{Croke06}, 
which optimizes the conditional probability that, when a given state is identified by the 
measurement, it is indeed the correct state.  
In another scheme considered in Refs. \cite{Chefles98, Zhang99, Fiurasek03, Eldar03}, 
the probability of correct discrimination is maximized while the rate of inconclusive results 
is fixed. This scheme interpolates in a way between minimum-error discrimination 
and unambiguous discrimination. 

In this paper we consider a discrimination problem with a finite error margin $m$ imposed on the 
probability of error. When the error margin $m$ is zero, the problem is equivalent to unambiguous 
discrimination. In minimum-error discrimination, no condition is imposed on the error probability, 
which means the error margin $m$ is 1.  Thus, discrimination with a general error margin 
unifies the two commonly adopted settings. 
A similar scheme was considered by Touzel, Adamson, and Steinberg \cite{Touzel07}, 
where they minimized the probability of an inconclusive result by imposing some bound on 
the error probability and compared the numerical results of projective and  
positive operator-valued measure (POVM) measurements.   
In Sec.\ref{sec:discrimination}, we will formulate two types of constraints on the error probability
and analytically determine the maximum success probability as a function of the error margin $m$ 
for two pure states with equal occurrence probabilities. 

Let us now assume that the two states to be discriminated are multipartite and generally 
entangled. The interesting question here is whether the parties sharing the input state 
can achieve the optimal discrimination by means of local operations and classical 
communication (LOCC).  It is known that this is possible for both minimum-error  
\cite{Walgate00,Virmani01} and unambiguous \cite{Chen02,Ji05} discrimination problems. 
In Sec.\ref{sec:locality}, we will tackle this problem for discrimination with a general 
error margin. By establishing a general theorem for three-element POVMs of a two-dimensional 
space, we show that discrimination with an error margin can be optimally performed by LOCC. 

\section{Discrimination with error margin of two pure states \label{sec:discrimination}}
We consider the problem of discriminating between two known pure states 
$\rho_1=\ket{\phi_1}{\bra{\phi_1}}$ and $\rho_2=\ket{\phi_2}{\bra{\phi_2}}$ with 
$|\braket{\phi_1}{\phi_2}| \ne 1$. 
We assume that the two states occur with equal probabilities and there is complete 
classical knowledge of the two states.    
Our measurement on the input state can produce three outcomes, $\mu=1,2,3$. 
The outcome $\mu=1$ or $2$ means that the input state is guessed to be $\rho_\mu$, 
and outcome 3 means that ``I do not know," which is called the inconclusive result. 
Let us introduce the POVM, $\{E_1,E_2,E_3\}$, 
corresponding to the three measurement outcomes.   

We define $P_{E_\mu,\rho_a}$ to be the joint probability that the input state is 
$\rho_a,(a=1,2)$ and the measurement outcome is $\mu=(1,2,3)$ produced by POVM element 
$E_\mu$:
\begin{eqnarray*}
   P_{E_\mu,\rho_a} = \frac{1}{2}\tr{E_\mu \rho_a}.
\end{eqnarray*}
The task is to maximize the discrimination success probability given by 
\begin{eqnarray}
 p_{\circ} = \sum_{a=1,2} P_{E_a,\rho_a} 
     = \frac{1}{2} \Big(  \tr{E_1 \rho_1} + \tr{E_2 \rho_2} \Big).  \label{eq:p_circle}
\end{eqnarray}

Let us define some probabilities of making errors. 
Suppose the measurement outcome is $\mu=1$. The probability of error in this case, 
which means that the input state was $\rho_2$, is the conditional probability defined by   
\begin{eqnarray}
 P_{\rho_2|E_1} = \frac{P_{E_1,\rho_2}}{P_{E_1}}, \label{eq:p_error_1} 
\end{eqnarray}
where $P_{E_1}=P_{E_1,\rho_1}+P_{E_1,\rho_2}$, the probability of 
finding outcome $\mu=1$. 
Similarly, when the measurement outcome is $\mu=2$, the conditional probability of 
error is given by
\begin{eqnarray}
 P_{\rho_1|E_2} = \frac{P_{E_2,\rho_1}}{P_{E_2}}. \label{eq:p_error_2}
\end{eqnarray}

In minimum-error discrimination, one maximizes the success probability of 
Eq.(\ref{eq:p_circle}) without imposing any conditions on the two conditional error probabilities 
of Eqs.(\ref{eq:p_error_1}) and (\ref{eq:p_error_2}).  
On the other hand, the conditional error probabilities of Eqs.(\ref{eq:p_error_1}) and (\ref{eq:p_error_2}) are required to be zero in the unambiguous discrimination problem. 

In this paper, we consider a discrimination problem with the conditions that these conditional 
error probabilities should not exceed a certain error margin $m\ (0 \le m \le 1)$,  
\begin{subequations}
   \label{eq:strong_margin}
\begin{eqnarray}
    P_{\rho_2|E_1} &=& \frac{\tr{E_1\rho_2}}{\tr{E_1\rho_1}+\tr{E_1\rho_2}} \le m, \\
    P_{\rho_1|E_2} &=& \frac{\tr{E_2\rho_1}}{\tr{E_2\rho_1}+\tr{E_2\rho_2}} \le m.                 
\end{eqnarray}
\end{subequations}
It is clear that unambiguous discrimination corresponds to the case of $m=0$ 
and minimum-error discrimination corresponds to the case of $m=1$ since the probabilities 
do not exceed 1. 

We can impose a margin of error in a different way.  
If we require that the mean error probability, denoted by $p_{\times}$, should not exceed the  
error margin $m$, we obtain   
\begin{eqnarray}
  p_{\times} \equiv P_{E_1,\rho_2}+P_{E_2,\rho_1} \le m. \label{eq:weak_margin}
\end{eqnarray}
It turns out that this condition is weaker than the error-margin conditions given in 
Eq.(\ref{eq:strong_margin}), because Eq.(\ref{eq:weak_margin}) follows from 
Eqs.(\ref{eq:strong_margin}):  
\begin{eqnarray*}
  p_{\times} &=& P_{E_1,\rho_2}+P_{E_2,\rho_1} 
   = P_{\rho_2|E_1}P_{E_1} + P_{\rho_1|E_2}P_{E_2}  \\ 
   & \le & m ( P_{E_1}+P_{E_2} ) \le m. 
\end{eqnarray*}
From now on, we call the conditions given by Eqs.(\ref{eq:strong_margin}) and 
Eq.(\ref{eq:weak_margin}) strong and weak error-margin conditions, respectively. 
We will first consider the discrimination problem with the strong error-margin condition. 
The weak error-margin condition will be discussed later. 

The discrimination problem with the strong error-margin condition is formulated 
in the following way:  
\begin{subequations} 
    \label{eq:sdp1}
\begin{eqnarray}
&&\mbox{maximize:} \nonumber \\
&&\ \ \ \ p_{\circ} = \frac{1}{2} \left(  \tr{E_1 \rho_1} + \tr{E_2 \rho_2} \right), \\
&&\mbox{subject to:} \nonumber \\
&&\ \ \ \ E_1 \ge 0,\ \ E_2 \ge 0,  \\ 
&&\ \ \ \ E_1+E_2 \le 1, \label{eq:E_3_positivity} \\
&&\ \ \ \ \tr{E_1\rho_2} \le m \left( \tr{E_1\rho_1} + \tr{E_1\rho_2} \right), 
                         \label{eq:strong_margin_1}  \\ 
&&\ \ \ \ \tr{E_2\rho_1} \le m \left( \tr{E_2\rho_1} + \tr{E_2\rho_2} \right).
                         \label{eq:strong_margin_2} 
\end{eqnarray}
\end{subequations}
Here, Eq.(\ref{eq:E_3_positivity}) represents the positivity condition of the POVM element $E_3$, 
and Eqs.(\ref{eq:strong_margin_1}) and (\ref{eq:strong_margin_2}) are the strong error-margin 
conditions. This is a problem of semidefinite programming. As we will see, this problem 
can be solved in a closed form. 

We present the results first. The maximum-discrimination success probability 
is given as follows: 
\begin{eqnarray}
  p_{\circ} = \left\{
  \begin{array}{ll}
     A_m\Big( 1 - |\braket{\phi_1}{\phi_2}| \Big)
               & (0 \le m \le m_c), \\
        \\
     \frac{1}{2}\left( 1 + \sqrt{1-|\braket{\phi_1}{\phi_2}|^2} \right) 
               & (m_c \le m \le 1), \\
  \end{array}
                  \right. \label{eq:p_circle_strong_max}
\end{eqnarray}
where
\begin{eqnarray}
  m_c=\frac{1}{2}\left( 1 - \sqrt{1-|\braket{\phi_1}{\phi_2}|^2} \right),  \label{eq:mc}
\end{eqnarray}
and $A_m$ is an increasing function of the error margin and defined to be 
\begin{eqnarray}
   A_m = \frac{1-m}{(1-2m)^2} \left( 1+2\sqrt{m(1-m)} \right).  \label{eq:am}
\end{eqnarray}

\begin{figure}
\includegraphics[width=8cm]{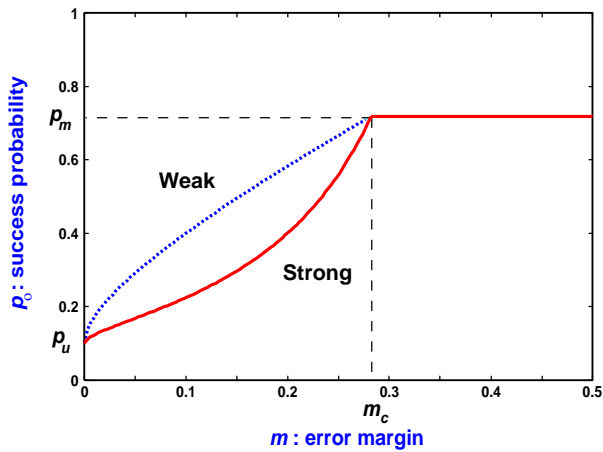}
\caption{\label{fig_strong_weak}
Discrimination success probability $p_{\circ}$. 
The solid line is the discrimination success probability with the strong error-margin 
condition. The dotted line is that with the weak error-margin condition. 
$p_u$ represents the success probability of unambiguous discrimination and 
$p_m$ is that of minimum-error discrimination. 
The fidelity $|\braket{\phi_1}{\phi_2}|$ is taken to be $0.9$. 
}
\end{figure}

Figure \ref{fig_strong_weak} displays the maximum success probability as a function of 
error margin. 
When the error margin is less than $m_c$, the maximum success probability is given by 
that of unambiguous discrimination, one minus the fidelity of the two states, multiplied by 
a factor $A_m$, which is an increasing function of the error margin. 
When $m=0$, the above $p_{\circ}$ reproduces the success probability of unambiguous 
discrimination since $A_0=1$.   
Note that the maximum success probability is equal to that of minimum-error discrimination 
when $m_c \le m \le 1$. This can be understood in the following way. In minimum-error 
discrimination, the success probability is optimized with no explicit conditions on errors. 
The resultant conditional error probabilities, $P_{\rho_2|E_1}$ and $P_{\rho_1|E_2}$, turn 
out to be $m_c$ given in Eq.(\ref{eq:mc}). 
Thus, an error margin greater than $m_c$ has no effect on the optimization of success probability.  
For $m_c \le m \le 1$, the optimal POVM is given by that of the minimum error discrimination 
problem.  

In what follows, we derive the maximum success probability of Eq.(\ref{eq:p_circle_strong_max}).
We work in the two-dimensional subspace spanned by states $\ket{\phi_1}$ and $\ket{\phi_2}$ 
so that the Bloch vector representation can be used: 
\begin{eqnarray*}
  \rho_a = \frac{1+\mbold{n}_a \cdot \mbold{\sigma}}{2},\ \ (a=1,2).
\end{eqnarray*}
We also parametrize POVM elements by Pauli matrices:
\begin{eqnarray*}
   E_\mu = \alpha_\mu + \mbold{\beta}_\mu \cdot \mbold{\sigma},\ \ (\mu=1,2,3).
\end{eqnarray*}
In terms of these parameters, the optimization problem takes the following form:
\begin{subequations}
  \label{eq:sdp2}
\begin{eqnarray}
&&\mbox{maximize:} \nonumber \\
&&\ \ \ \ p_{\circ} = \frac{1}{2} \left( \alpha_1 + \mbold{\beta}_1 \cdot \mbold{n}_1 
                + \alpha_2 + \mbold{\beta}_2 \cdot \mbold{n}_2  \right), \\
&&\mbox{subject to:} \nonumber \\
&&\ \ \ \ \alpha_1 \ge |\mbold{\beta}_1|,\ \ \alpha_2 \ge |\mbold{\beta}_2|, \label{eq:e12} \\ 
&&\ \ \ \ \alpha_1+\alpha_2+|\mbold{\beta}_1+\mbold{\beta}_2| \le 1, \label{eq:e3} \\
&&\ \ \ \ \alpha_1+\mbold{\beta}_1 \cdot \mbold{n}_2 
            \le m \left( 2\alpha_1 + \mbold{\beta}_1 \cdot (\mbold{n}_1+\mbold{n}_2) \right), 
                      \label{eq:m1}  \\ 
&&\ \ \ \ \alpha_2+\mbold{\beta}_2 \cdot \mbold{n}_1 
            \le m \left( 2\alpha_2 + \mbold{\beta}_2 \cdot (\mbold{n}_1+\mbold{n}_2) \right).
                      \label{eq:m2}
\end{eqnarray}
\end{subequations}
The variables in this optimization problem are the parameters 
$\{\alpha_1,\mbold{\beta}_1,\alpha_2,\mbold{\beta}_2\}$. 
Note that the conditions (\ref{eq:e12}), (\ref{eq:e3}), (\ref{eq:m1}), and (\ref{eq:m2}) define 
a convex set in the parameter space; if each parameter set of   $\{\alpha_1,\mbold{\beta}_1,\alpha_2,\mbold{\beta}_2\}$ 
and $\{\alpha_1',\mbold{\beta}_1',\alpha_2',\mbold{\beta}_2'\}$ 
satisfies the conditions, so does the set 
\begin{eqnarray*}
 \{p\alpha_1+q\alpha_1',\,p\mbold{\beta}_1+q\mbold{\beta}_1',\,p\alpha_2+q\alpha_2',\,
   p\mbold{\beta}_2+q\mbold{\beta}_2'\},
\end{eqnarray*}
for all $p,q \ge 0$ with $p+q=1$. 
This convexity enables us to impose two useful symmetries on optimal 
parameters without loss of generality. 

We consider the symmetry with respect to exchange of the two Bloch vectors 
$\mbold{n}_1$ and $\mbold{n}_2$. Let $O$ be the reflection matrix with respect to the 
plane which is perpendicular to $\mbold{n}_1-\mbold{n}_2$ and contains the origin 
of the Bloch sphere, so that $O\mbold{n}_1=\mbold{n}_2$ and $O\mbold{n}_2=\mbold{n}_1$. 
Suppose the parameter set $\{\alpha_1,\mbold{\beta}_1,\alpha_2,\mbold{\beta}_2\}$ is optimal. 
It is easy to see that the parameter set $\{\alpha_2,O\mbold{\beta}_2,\alpha_1,O\mbold{\beta}_1\}$ 
is optimal. Furthermore, by convexity, the arithmetic average of these two sets of parameters 
is also optimal since the success probability $p_{\circ}$ is linear in the parameters.
The parameter set obtained in this way clearly satisfies the following symmetry:
\begin{eqnarray}
  \alpha_1=\alpha_2,\ \ \mbold{\beta}_2=O\mbold{\beta}_1.  \label{eq:reflexion}
\end{eqnarray}
We can repeat a similar argument for the reflection with respect to the plane 
that contains the vectors $\mbold{n}_1$ and $\mbold{n}_2$. This consideration enables us to 
safely assume that the vectors $\mbold{\beta}_1$ and  $\mbold{\beta}_2$ lie on this plane. 

Let us now show that equality should actually hold in the inequality conditions 
(\ref{eq:e12}) and (\ref{eq:e3}) for optimal parameters.  
This implies that the rank of each element of the POVM $\{E_1,E_2,E_3\}$ 
does not exceed 1; this property will be important in the next section. 
We begin with the condition (\ref{eq:e3}) and assume that equality does not hold 
for an optimal set of parameters. Then we can multiply all parameters by a common positive number 
greater than one so that all the conditions (\ref{eq:e12}), (\ref{eq:e3}), (\ref{eq:m1}), 
and (\ref{eq:m2}) are still satisfied while the success probability $p_\circ$ increases.  
Since this is a contradiction, we can conclude that equality holds in condition (\ref{eq:e3}).

A similar, but rather involved, argument can also be applied to condition (\ref{eq:e12}). 
Assume that a strict inequality holds in Eq.(\ref{eq:e12}) for an optimal set of parameters 
with the symmetry given by Eq.(\ref{eq:reflexion}). Let us increase the component 
of $\mbold{\beta}_1$ in the direction of $(\mbold{n}_1-\mbold{n}_2)$ while leaving the component 
along vector $(\mbold{n}_1+\mbold{n}_2)$ unchanged. Note that $\mbold{\beta}_2$ also changes 
in accordance with the symmetry of (\ref{eq:reflexion}).  
The left-hand side of (\ref{eq:e3}) and the right-hand sides of (\ref{eq:m1}) and (\ref{eq:m2}) 
remain the same. The left-hand sides of (\ref{eq:m1}) and (\ref{eq:m2}) decrease, while 
the success probability $p_{\circ}$ increases.  
It is clear that there exists a positive increment of the component of $\mbold{\beta}_1$ 
along the direction $(\mbold{n}_1-\mbold{n}_2)$ 
such that $p_{\circ}$ increases while all the conditions are still satisfied. 
We thus conclude that equality holds in condition (\ref{eq:e12}). 

From these considerations, we can rewrite the problem in terms of variables 
$\alpha \equiv \alpha_1$ and $\mbold{\beta} \equiv \mbold{\beta}_1$:
\begin{eqnarray*}
&&\mbox{maximize:} \nonumber \\
&&\ \ \ \ p_{\circ} = \alpha + \mbold{\beta} \cdot \mbold{n}_1,  \\
&&\mbox{subject to:} \nonumber \\
&&\ \ \ \ \alpha = |\mbold{\beta}|, \\ 
&&\ \ \ \ 2\alpha+|(1+O)\mbold{\beta}| =1 ,  \\
&&\ \ \ \ \alpha + \mbold{\beta} \cdot \mbold{n}_2 
            \le m \left( 2\alpha + \mbold{\beta}\cdot (\mbold{n}_1+\mbold{n}_2) \right).
\end{eqnarray*}
Since the vector $\mbold{\beta}$ lies on the plane spanned by $\mbold{n}_1$ and $\mbold{n}_2$, 
we can expand $\mbold{\beta}$ as 
\begin{eqnarray*}
  \mbold{\beta} = x \frac{\mbold{n}_1+\mbold{n}_2}{2} + y \frac{\mbold{n}_1-\mbold{n}_2}{2}.
\end{eqnarray*}
The problem is further simplified in terms of variables $\alpha$, $x$, $y$ and takes 
the following form:
\begin{subequations}
  \label{eq:sdp3}
\begin{eqnarray}
&&\mbox{maximize:} \nonumber \\
&&\ \ \ \ p_{\circ} = \alpha + Sx + Ty, \label{eq:Pcirc} \\
&&\mbox{subject to:} \nonumber \\
&&\ \ \ \ \alpha = \sqrt{Sx^2 + Ty^2}, \label{eq:E1} \\ 
&&\ \ \ \ 2\alpha+2\sqrt{S}|x| =1 , \label{eq:E3} \\
&&\ \ \ \ (1-2m)(\alpha+Sx) -Ty \le 0.  \label{eq:M}
\end{eqnarray}
\end{subequations} 
Here, we introduced positive constants $S$ and $T$:
\begin{eqnarray*}
  S & \equiv & \frac{1+\mbold{n}_1\cdot \mbold{n}_2}{2} = |\braket{\phi_1}{\phi_2}|^2 , \\
  T & \equiv & \frac{1-\mbold{n}_1\cdot \mbold{n}_2}{2} = 1-|\braket{\phi_1}{\phi_2}|^2 .
\end{eqnarray*}
By using Eqs.(\ref{eq:E1}) and (\ref{eq:E3}), we can express $\alpha$ and $x$ in terms of $y$:
\begin{eqnarray*}
  \alpha &=& Ty^2 + \frac{1}{4}, \\
  \sqrt{S}|x| &=& \frac{1}{4}(1-4Ty^2). 
\end{eqnarray*}
Condition (\ref{eq:M}) is then a quadratic inequality of variable $y$. 
Optimization can now be explicitly performed since the success probability 
$p_\circ$ becomes a quadratic function of $y$. 
After a long tedious calculation, which is outlined in the appendix, 
the maximum success probability 
is found to be given by Eq.(\ref{eq:p_circle_strong_max}). 
When the error margin is in the range of $0 \le m \le m_c$, the optimal POVM 
are given by the parameters 
$\alpha_{\max}$, $x_{\max}$, and $y_{\max}$: 
\begin{eqnarray}
  y_{\max} & \equiv & \frac{1+2\sqrt{m(1-m)}}{2(1+\sqrt{S})(1-2m)}, \label{eq:ymax}\\
  x_{\max} & \equiv & -\frac{1}{4\sqrt{S}}(1-4Ty_{\max}^2), \label{eq:xmax} \\
  \alpha_{\max} & \equiv & Ty_{\max}^2 + \frac{1}{4}. \label{eq:alphamax}
\end{eqnarray}

Before concluding this section, we briefly summarize the results of the weak error-margin condition 
given in Eq.(\ref{eq:weak_margin}). For the weak error-margin condition, 
conditions (\ref{eq:m1}) and (\ref{eq:m2}) should be replaced by 
\begin{eqnarray*}
  \alpha_1 + \mbold{\beta}_1 \cdot \mbold{n}_2 + \alpha_2 + \mbold{\beta}_2 \cdot \mbold{n}_1 \le 2m.  
\end{eqnarray*}
We can proceed in a similar way to the strong error-margin case. 
Omitting the details of derivation, we present the maximum success probability with the 
weak error-margin condition:  
\begin{eqnarray}
  p_{\circ} = \left\{
  \begin{array}{ll}
     \Big( \sqrt{m} + \sqrt{1 - |\braket{\phi_1}{\phi_2}| } \Big)^2
               & (0 \le m \le m_c),  \\
        \\
     \frac{1}{2}\left( 1 + \sqrt{1-|\braket{\phi_1}{\phi_2}|^2} \right) 
               & (m_c \le m \le 1). \\
  \end{array}
                  \right. \label{eq:p_circle_weak_max}
\end{eqnarray}
Unlike the strong error-margin case, the success probability for $0 \le m \le m_c$ 
is not simply proportional to that of unambiguous discrimination. 
For purpose of comparison, the success probabilities of the two error-margin 
conditions are plotted in Fig. \ref{fig_strong_weak}. We note that each element of the 
optimal POVM in this case is again of rank 0 or 1, though it is more involved to show 
than in the strong error-margin case. The optimal POVM for $0 \le m \le m_c$ is given by 
the parameters: 
\begin{eqnarray*}
  y_{\max} & \equiv & \frac{1}{2(1+\sqrt{S})}\left( 1+2\sqrt{\frac{m}{1-\sqrt{S}}} \right), \\
  x_{\max} & \equiv & -\frac{1}{4\sqrt{S}}(1-4Ty_{\max}^2), \\
  \alpha_{\max} & \equiv & Ty_{\max}^2 + \frac{1}{4}.
\end{eqnarray*}

\section{Local discrimination with error margin \label{sec:locality}}
Suppose that the two pure states to be discriminated are multipartite and generally 
entangled. Can the parties sharing the input state perform optimal discrimination 
by LOCC? It has been shown that this is possible in both the minimum-error 
\cite{Walgate00,Virmani01} and unambiguous \cite{Chen02,Ji05} discrimination problems. 
As shown in the preceding section, those correspond to the cases of $m_c \le m \le 1$ 
and $m=0$. In this section we will show that this is true for any value of error margin. 

In the preceding section we determined the optimal POVM in the two-dimensional 
subspace spanned by $\ket{\phi_1}$ and $\ket{\phi_2}$ and showed that the rank 
of each POVM element does not exceed 1 for any error margin.  
We will show that the following general theorem holds. 

\bigskip\noindent{\bf Theorem 1}: {\it 
Let $V$ be a two-dimensional subspace of a multipartite tensor-product space $H$, and $P$ be 
the projector onto the subspace $V$. Then, for any three-element POVM 
$\{E_1,E_2,E_3\}$ of $V$ with every element being of rank 0 or 1, there exists a one-way LOCC POVM 
$\{E_1^{{\rm L}},E_2^{{\rm L}},E_3^{{\rm L}}\}$ of $H$ such that  
$
  E_\mu = P E_\mu^{{\rm L}} P \ (\mu=1,2,3). 
$
}\bigskip

This implies that a POVM satisfying the conditions of Theorem 1 can be implemented 
by a one-way LOCC protocol as far as measurement for states in subspace $V$ is  
concerned. The optimal POVM of discrimination with error margin satisfies the 
conditions of Theorem 1; therefore, it is achievable by a one-way LOCC protocol. 

In the rest of the section, we prove Theorem 1. To do so, it suffices to develop the proof 
of Ji et al., which shows that unambiguous discrimination of two pure 
states with arbitrary occurrence probabilities can be optimally realized by means of 
a one-way LOCC \cite{Ji05}. 
Our strategy for the proof is the following. 
First, we show that the optimal POVMs of the global and the LOCC schemes share the 
same matrix elements for all states in $V$, which means that Theorem 1 holds for 
the optimal POVM of any unambiguous discrimination problem. 
Then, we show that any POVM satisfying the conditions of Theorem 1 can be 
regarded as the optimal POVM of a certain unambiguous discrimination problem. 

Let us consider unambiguous discrimination between pure 
states $\ket{\Phi_1}$ and $\ket{\Phi_2}$ with occurrence probabilities $s$ and $t$, 
respectively. We assume $|\braket{\Phi_1}{\Phi_2}| \ne 1$ and denote by $V$ the 
two-dimensional subspace spanned by $\ket{\Phi_1}$ and $\ket{\Phi_2}$. 
We begin with the global discrimination scheme. 
For our purpose, it suffices to consider the case where the following conditions are 
satisfied:
\begin{eqnarray*}
  \sqrt{\frac{s}{t}},\ \sqrt{\frac{t}{s}} \ge |\braket{\Phi_1}{\Phi_2}|. 
\end{eqnarray*}
In this case the optimal POVM is given by 
\begin{eqnarray}
\left\{ 
  \begin{array}{lll}
    E_1 &=& a_1 \ket{\Phi_2^\perp}\bra{\Phi_2^\perp}, \\
    E_2 &=& a_2 \ket{\Phi_1^\perp}\bra{\Phi_1^\perp}, \\
    E_3 &=& P-E_1-E_2. \\
  \end{array}
\right.
  \label{eq:global_POVM}
\end{eqnarray}
Here, $\ket{\Phi_a^\perp}\ (a=1,2)$ is a normalized state in $V$ which is orthogonal to $\ket{\Phi_a}$. 
The $\ket{\Phi_a^\perp}$ is unique up to a phase factor. 
The coefficients $a_1$ and $a_2$ are given by 
\begin{subequations}
   \label{eq:global_a}
\begin{eqnarray}
    a_1 &=& \frac{1-\sqrt{\frac{t}{s}}\,|\braket{\Phi_1}{\Phi_2}|}{1-|\braket{\Phi_1}{\Phi_2}|^2}, 
              \\
    a_2 &=& \frac{1-\sqrt{\frac{s}{t}}\,|\braket{\Phi_1}{\Phi_2}|}{1-|\braket{\Phi_1}{\Phi_2}|^2},
\end{eqnarray}
\end{subequations}
which implies that $E_3$ is also of rank 0 or 1.

Now suppose that the states $\ket{\Phi_1}_{AB}$ and $\ket{\Phi_2}_{AB}$ are bipartite, shared by 
Alice and Bob. We can choose an appropriate phase for the states so that 
$\braket{\Phi_1}{\Phi_2}$ has a nonnegative real value. 
Ji et al. showed that, by appending an ancillary system $R$ to, say, Alice's system $A$, we 
can choose an orthonormal basis $\{\ket{I}_{RA}\}$ for Alice's combined system $RA$ so that 
the following relations hold: 
\begin{eqnarray}
  \ket{0}_R \otimes \ket{\Phi_1}_{AB} &=& \sum_I \sqrt{s_I}\ket{I}_{RA} \otimes \ket{\eta_I}_B, \\
  \ket{0}_R \otimes \ket{\Phi_2}_{AB} &=& \sum_I \sqrt{t_I}\ket{I}_{RA} \otimes \ket{\gamma_I}_B,
\end{eqnarray}
where $\ket{\eta_I}$ and $\ket{\gamma_I}$ are normalized states of Bob's system $B$, 
and $\braket{\eta_I}{\gamma_I}$ is a nonnegative real number satisfying 
\begin{eqnarray}
  \sqrt{\frac{ss_I}{tt_I}},\ \sqrt{\frac{tt_I}{ss_I}} \ge \braket{\eta_I}{\gamma_I} \ge 0.
\end{eqnarray}
This decomposition of the states defines a one-way LOCC protocol: Alice first performs 
measurement in the basis $\{\ket{I}\}$ and informs Bob of the outcome $I$; he then 
discriminates between states $\ket{\eta_I}$ and $\ket{\gamma_I}$. Ji et al. showed that this 
one-way LOCC protocol achieves the maximum success probability given by the global optimal 
POVM of Eq.(\ref{eq:global_POVM}). 
 
We can show that the POVM $\{E_\mu^{{\rm L}}\}_{\mu=1,2,3}$ corresponding to the protocol of 
Ji et al. actually satisfies stronger conditions; the global POVM and the LOCC POVM share 
the same matrix element between any states in the subspace $V$.  To see this, let us 
write the LOCC POVM $\{E_\mu^{{\rm L}}\}_{\mu=1,2,3}$ as  
\begin{eqnarray}
  E_\mu^{{\rm L}} = \sum_I e_I^{A} \otimes e_\mu^{B}(I)\ \ (\mu = 1,2,3),  \label{eq:LOCC_POVM}
\end{eqnarray}
where $\{e_I^{A}\}_I$ is a POVM of Alice's system $A$ defined by 
\begin{eqnarray*}
   e_I^{A} = {}_R\braket{0}{I}_{(RA)}\braket{I}{0}_R,  
\end{eqnarray*}
and $\{e_\mu^{B}(I)\}_{\mu=1,2,3}$ is Bob's POVM: 
\begin{subequations}
   \label{eq:Bob_POVM}
\begin{eqnarray}
   e_1^{B}(I) &=&  \frac{1-\sqrt{\frac{tt_I}{ss_I}}\,\braket{\eta_I}{\gamma_I}}
                        {1-\braket{\eta_I}{\gamma_I}^2}
                        \,\ket{\gamma_I^\perp}\bra{\gamma_I^\perp},  \\
   e_2^{B}(I) &=&  \frac{1-\sqrt{\frac{ss_I}{tt_I}}\,\braket{\eta_I}{\gamma_I}}
                        {1-\braket{\eta_I}{\gamma_I}^2}
                        \,\ket{\eta_I^\perp}\bra{\eta_I^\perp}, \\
   e_3^{B}(I) &=&  P^{B}(I) -  e_1^{B}(I) - e_2^{B}(I). 
\end{eqnarray}
\end{subequations}
Here, operator $P^{B}(I)$ is the projector onto the two-dimensional subspace $V^{B}(I)$ spanned by 
$\ket{\eta_I}$ and $\ket{\gamma_I}$, and the state $\ket{\eta_I^\perp}$ is orthogonal to 
$\ket{\eta_I}$ in this subspace. $\ket{\gamma_I^\perp}$ is defined similarly. 

The global $E_1$ has the following vanishing matrix elements: 
\begin{eqnarray*}
  \bra{\Phi_2}E_1\ket{\Phi_2}=\bra{\Phi_2}E_1\ket{\Phi_1}=\bra{\Phi_1}E_1\ket{\Phi_2}=0.  
\end{eqnarray*}
It is clear that the corresponding matrix elements of $E_1^{{\rm L}}$ are also zero. 
As for the matrix element between $\ket{\Phi_1}$ and $\ket{\Phi_1}$, $E_1$ gives 
\begin{eqnarray*}
  \bra{\Phi_1}E_1\ket{\Phi_1}=a_1|\braket{\Phi_1}{\Phi_2^\perp}|^2
          =1-\sqrt{\frac{t}{s}}\braket{\Phi_1}{\Phi_2}. 
\end{eqnarray*}
We find that $E_1^{{\rm L}}$ has the same matrix element: 
\begin{eqnarray*}
  \bra{\Phi_1}E_1^{{\rm L}}\ket{\Phi_1} &=& 
   \sum_I s_I \frac{1-\sqrt{\frac{tt_I}{ss_I}}\,\braket{\eta_I}{\gamma_I}}{1-\braket{\eta_I}{\gamma_I}^2}
   |\braket{\eta_I}{\gamma_I^\perp}|^2  \\
      &=& 1-\sqrt{\frac{t}{s}}\braket{\Phi_1}{\Phi_2}. 
\end{eqnarray*}
The same thing can be readily verified for $E_2^{{\rm L}}$ and $E_3^{{\rm L}}$. 
We thus have shown that $E_\mu = P E_\mu^{{\rm L}} P,\ (\mu=1,2,3)$.

Now let us consider a general POVM $\{E_\mu\}_{\mu=1,2,3}$ that satisfies the conditions of 
Theorem 1, and write it in terms of normalized states $\ket{\psi_\mu}$ and nonnegative coefficients 
$b_\mu$: 
\begin{eqnarray}
    E_\mu &=& b_\mu \ket{\psi_\mu}\bra{\psi_\mu}\ \ (\mu=1,2,3).  \label{eq:generalPOVM}
\end{eqnarray}
There are two linearly independent states among $\{\ket{\psi_\mu}\}_{\mu=1,2,3}$, which 
can be assumed to be $\ket{\psi_1}$ and $\ket{\psi_2}$. The two-dimensional subspace $V$ is 
spanned by those states. Operator $E_3=P-E_1-E_2$ is of rank 0 or 1 by definition; therefore,  
the greater eigenvalue of the two eigenvalues of operator $E_1+E_2$ is equal to 1. 
This implies that the coefficients $b_1$ and $b_2$ can be expressed as 
\begin{subequations}
\begin{eqnarray}
  b_1 &=& \frac{1-r\,|\braket{\psi_1}{\psi_2}|}{1-|\braket{\psi_1}{\psi_2}|^2}, 
                  \label{eq:b_1} \\
  b_2 &=& \frac{1-\frac{1}{r}\,|\braket{\psi_1}{\psi_2}|}{1-|\braket{\psi_1}{\psi_2}|^2},
                  \label{eq:b_2}
\end{eqnarray}
\end{subequations}
where $r$ is a positive number that satisfies the condition $r,\ 1/r \ge |\braket{\psi_1}{\psi_2}|$. 
This can be seen in the following way. It is evident that there exists a positive number $r$ 
such that the coefficient $b_1$ can be written in the form of Eq.(\ref{eq:b_1}) 
since $0 \le b_1 \le 1$. It is also easy to see that if the greater eigenvalue of $E_1+E_2$ 
is equal to 1, then the coefficients $b_1$ and $b_2$ should satisfy the relation:  
\begin{eqnarray*}
  b_1 + b_2 = 1 + b_1 b_2 \left( 1 - |\braket{\psi_1}{\psi_2}|^2 \right).  
\end{eqnarray*}
From this relation we find that $b_2$ is expressed by Eq.(\ref{eq:b_2}). 

For any positive $r$, there is a set of positive numbers $s$ and $t$ such that 
$r=\sqrt{t/s}$ and $s+t=1$. We also note that the relation 
$|\braket{\psi_1}{\psi_2}|=|\braket{\psi_1^\perp}{\psi_2^\perp}|$ holds.  
Therefore, a general POVM (\ref{eq:generalPOVM}) satisfying the conditions of Theorem 1 can be 
identified with the optimal POVM of unambiguous discrimination between 
$\ket{\psi_2^\perp}$ and $\ket{\psi_1^\perp}$ with the occurrence probabilities $s$ and $t$, 
respectively. Thus, we conclude that Theorem 1 holds for the bipartite case.    

The general multipartite case of Theorem 1 follows by induction. 
Remember that Bob's POVM is given by Eq.(\ref{eq:Bob_POVM}), which satisfies 
the conditions of Theorem 1: this is a POVM of the two-dimensional subspace $V^{B}(I)$ and 
each element is of rank 0 or 1. If Bob's system is composite and actually shared by Bob 
and Charles, we can construct a one-way LOCC POVM for Bob and Charles in the same way. 
The whole one-way LOCC POVM would take the form 
\begin{eqnarray*}
    E_\mu^{{\rm L}} = \sum_{I_AI_B} e_{I_A}^{A} \otimes e_{I_B}^{B}(I_A) 
                      \otimes e_\mu^{C}(I_A,I_B)\ \ (\mu = 1,2,3). 
\end{eqnarray*}
It is clear that we can repeat the same procedure if necessary. 
This completes the proof of Theorem 1. 

\section{Concluding Remarks}
We considered a quantum-state discrimination problem with an error margin imposed, 
which unifies the minimum-error and the unambiguous discrimination problems. 
We determined the optimal discrimination success probability for two pure states 
with equal occurrence probabilities.  When the states are multipartite, 
we have shown that a LOCC scheme achieves the globally attainable optimal success 
probability. This was shown by establishing a general theorem for three-element POVMs
of a two-dimensional space. 

In this paper, we assumed that complete classical knowledge is given for the states to be 
discriminated. Instead, we can consider a situation where no classical knowledge of the states 
is given, but a certain number of their copies are available as reference states. 
One's task is to correctly identify a given input state with one of the reference states 
by some measurement on the whole state \cite{Hayashi05,Bergou05,Hayashi06}. 
This problem is called the quantum-state identification problem to distinguish it from 
discrimination problems with classical knowledge assumed. 
We studied the identification problem for two bipartite pure states 
\cite{Ishida08} and found that the optimal unambiguous identification 
cannot be achieved by LOCC while this is possible in minimum-error identification. 
It would be of great interest to consider identification problems with a general error 
margin imposed.

\appendix*
\section{}
Here we outline the last part of the derivation for the optimal success probability 
given by Eq.(\ref{eq:p_circle_strong_max}). We begin with the optimization problem: 
\begin{subequations}
\begin{eqnarray}
&&\mbox{maximize:} \nonumber \\
&&\ \ \ \ p_{\circ} = \alpha + Sx + Ty, \label{eq:A_opjective} \\
&&\mbox{subject to:} \nonumber \\
&&\ \ \ \ (1-2m)(\alpha+Sx) -Ty \le 0,  \label{eq:A_constraint} \\
&&\mbox{where} \nonumber \\
&&\ \ \ \ \alpha = Ty^2 + \frac{1}{4}, \label{eq:A_alpha} \\
&&\ \ \ \ \sqrt{S}|x| = \frac{1}{4}(1-4Ty^2). \label{eq:A_x} 
\end{eqnarray}
\end{subequations} 
We consider only the case of $0 \le m < m_c \equiv (1-\sqrt{T})/2$, 
since we already know that the minimum-error strategy is optimal when 
$m_c \le m \le 1$.  By our assumption $|\braket{\phi_1}{\phi_2}| \ne 1$, 
we have $S \ne 1$, $T \ne 0$, and $m_c < 1/2$. 
Note that Eq.(\ref{eq:A_x}) implies $|y| \le 1/(2\sqrt{T})$ 
and also shows that it is convenient to treat separately the two cases of different signs 
of $x$.

First we will show that there is no feasible solution if $x \ge 0$.  
In this case, the condition of Eq.({\ref{eq:A_constraint}) means that 
the quadratic function of $y$ defined by 
\begin{eqnarray*}
 f(y) \equiv T(1-\sqrt{S})y^2 -\frac{T}{1-2m}y + \frac{1+\sqrt{S}}{4},
\end{eqnarray*}
should be nonpositive.  The function $f(y)$ has the vertex at
\begin{eqnarray*} 
 y=\frac{1}{2(1-2m)(1-\sqrt{S})} \ge \frac{1}{2\sqrt{T}}.
\end{eqnarray*}
On the other hand, we find  
\begin{eqnarray*}
  f\left(\frac{1}{2\sqrt{T}}\right) = \frac{1}{2} \left( 1- \frac{\sqrt{T}}{1-2m} \right) > 0, 
\end{eqnarray*}
which implies $f(y) > 0$ for $y \le 1/(2\sqrt{T})$.
Thus the case of $x \ge 0$ is not feasible. 

We now assume $x < 0$. The success probability is then given by 
\begin{eqnarray*}
 p_{\circ}(y) \equiv T(1+\sqrt{S})y^2 + Ty + \frac{1-\sqrt{S}}{4}.  \label{eq:A_opjective2}
\end{eqnarray*}
The condition of Eq.({\ref{eq:A_constraint}) implies $g(y) \le 0$, where 
the quadratic function $g(y)$ is defined to be 
\begin{eqnarray*}
  g(y) \equiv T(1+\sqrt{S})y^2 -\frac{T}{1-2m}y + \frac{1-\sqrt{S}}{4}. 
\end{eqnarray*}
The function $g(y)$ has the vertex at 
\begin{eqnarray*}
  y=y_0 \equiv \frac{1}{2(1-2m)(1+\sqrt{S})} < \frac{1}{2\sqrt{T}}, 
\end{eqnarray*}
and we also find 
\begin{eqnarray*}
  g(y_0) = \frac{T}{4(1+\sqrt{S})}\left(1-\frac{1}{(1-2m)^2} \right) \le 0.
\end{eqnarray*}
On the other hand, we have 
\begin{eqnarray*}
  g\left(\frac{1}{2\sqrt{T}}\right) = \frac{1}{2} \left( 1- \frac{\sqrt{T}}{1-2m} \right) > 0. 
\end{eqnarray*}
Since $p_{\circ}(y)$ is an increasing function for 
positive $y$, the optimal $p_{\circ}$ is given by the greater root of the quadratic 
equation $g(y)=0$, which is given by   
\begin{eqnarray*}
  y_{\max} = \frac{1+2\sqrt{m(1-m)} }{2(1+\sqrt{S})(1-2m)}.
\end{eqnarray*} 
Substituting $y_{\max}$ in $p_{\circ}(y)$, we obtain the optimal success probability 
\begin{eqnarray*}
  p_{\circ}(y_{\max}) &=& g(y_{\max})+\frac{T}{1-2m}y_{\max}+Ty_{\max} \\
                      &=& A_m (1-\sqrt{S}), 
\end{eqnarray*}
where $A_m$ is defined by Eq.(\ref{eq:am}). It is clear that the other optimal parameters 
$x$ and $\alpha$ are given by Eq.(\ref{eq:xmax}) and Eq.(\ref{eq:alphamax}), respectively.


\begin{thebibliography}{99}
\bibitem{Chefles00}
A.~Chefles, 
Contemp. Phys. {\bf 41}, 401 (2000). 

\bibitem{Helstrom76}
C.~W.~Helstrom,
{\it Quantum Detection and Estimation Theory} 
(Academic Press, New York, 1976).

\bibitem{Ivanovic87}
I.~D.~Ivanovic,
Phys. Lett. A {\bf 123}, 257 (1987).

\bibitem{Dieks88}
D.~Dieks,
Phys. Lett. A {\bf 126}, 303 (1988).

\bibitem{Peres88} 
A.~Peres,
Phys. Lett. A {\bf 128}, 19 (1988). 

\bibitem{Jaeger95}
G.~Jaeger and A.~Shimony, 
Phys. Lett. A {\bf 197}, 83 (1995). 

\bibitem{Croke06}
Sarah~Croke, Erika~Andersson, Stephen~M.~Barnett, 
Claire~R.~Gilson, and John~Jeffers,  
Phys. Rev. Lett. {\bf 96}, 070401 (2006).

\bibitem{Chefles98}
A.~Chefles and S.~M.~Barnett, 
J. Mod. Opt. {\bf 45}, 1295 (1998). 

\bibitem{Zhang99}
Chuan-Wei Zhang, Chuan-Feng Li, and Guang-Can Guo,  
Phys. Lett. A {\bf 261}, 25 (1999).

\bibitem{Fiurasek03}
J.~Fiurasek and M.~Jezek, 
Phys. Rev. A {\bf 67}, 012321 (2003). 

\bibitem{Eldar03}
Y.~C.~Eldar, 
Phys. Rev. A {\bf 67}, 042309 (2003).  

\bibitem{Touzel07}
M.~A.~P.~Touzel, R.~B.~A.~Adamson, and A.~M.~Steinberg
Phys. Rev. A {\bf76}, 062314 (2007). 

\bibitem{Walgate00}
Jonathan~Walgate, Anthony~J.~Short, Lucien~Hardy, and Vlatko~Vedral,
Phys. Rev. Lett. {\bf 85}, 4972 (2000).

\bibitem{Virmani01}
S.~Virmani, M.~F.~Sacchi, M.~B.~Plenio, and D.~Markham, 
Phys. Lett. A {\bf 288}, 62 (2001). 

\bibitem{Chen02}
Y.-X.~Chen and D.~Yang, 
Phys. Rev. A {\bf 65}, 022320 (2002).

\bibitem{Ji05} 
Zhengfeng~Ji, Hongen~Cao, and Mingsheng~Ying,
Phys. Rev. A {\bf 71}, 032323 (2005). 

\bibitem{Hayashi05}
A.~Hayashi, M.~Horibe, and T.~Hashimoto, 
Phys. Rev. A {\bf 72}, 052306 (2005). 

\bibitem{Bergou05}
Janos~A.~Bergou and Mark~Hillery,
Phys. Rev. Lett. {\bf 94}, 160501 (2005).

\bibitem{Hayashi06}
A.~Hayashi, M.~Horibe, and T.~Hashimoto, 
Phys. Rev. A {\bf 73}, 012328 (2006). 

\bibitem{Ishida08}
Y.~Ishida, T.~Hashimoto, and M.~Horibe, and A.~Hayashi,  
Phys. Rev. A {\bf 78}, 012309 (2008).

\end{thebibliography}
\end{document}